\documentclass[fleqn,10pt]{olplainarticle}

\usepackage[hidelinks]{hyperref}
\usepackage{comment}
\usepackage{color}
\usepackage{booktabs} 
\usepackage{array}  

\title{Software Sustainability \& High Energy Physics}

\author[1]{Daniel S. Katz}
\affil[1]{d.katz@ieee.org, University of Illinois at Urbana-Champaign, USA, 0000-0001-5934-7525}
\author[2]{Sudhir Malik}
\affil[2]{sudhir.malik@upr.edu, University of Puerto Rico Mayaguez, USA, 0000-0002-6356-2655}
\author[3]{Mark S. Neubauer}
\affil[3]{msn@illinois.edu, University of Illinois at Urbana-Champaign, USA, 0000-0001-8434-9274}
\author[4]{Graeme A. Stewart}
\affil[4]{graeme.andrew.stewart@cern.ch, CERN, Switzerland, 0000-0003-0182-7088}
\author[5]{K\'et\'evi A. Assamagan}
\affil[5]{ketevi@bnl.gov, Brookhaven National Laboratory, USA, 0000-0002-4826-2662}
\author[6]{Erin A. Becker}
\affil[6]{ebecker@carpentries.org, The Carpentries, USA, 0000-0002-6832-0233}
\author[7]{Neil P. Chue Hong}
\affil[7]{N.ChueHong@software.ac.uk, University of Edinburgh, UK, 0000-0002-8876-7606}
\author[8]{Ian A. Cosden}
\affil[8]{icosden@princeton.edu, Princeton University, USA, 0000-0003-3780-9172}
\author[9]{Samuel Meehan}
\affil[9]{samuel.meehan@cern.ch, CERN, Switzerland, 0000-0002-3613-7514}
\author[10]{Edward J. W. Moyse}
\affil[10]{edward.moyse@cern.ch, University of Massachusetts, Amherst, USA, 0000-0003-4449-6178}
\author[11]{Adrian M. Price-Whelan}
\affil[11]{adrianmpw@gmail.com, Center for Computational Astrophysics, Flatiron Institute, USA, 0000-0003-0872-7098}
\author[12]{Elizabeth Sexton-Kennedy}
\affil[12]{sexton@fnal.gov, Fermilab, USA, 0000-0001-9171-1980}
\author[13]{Meirin Oan Evans}
\affil[13]{meirin.oan.evans@cern.ch, University of Sussex, Brighton, UK, 0000-0002-4259-018X}
\author[14]{Matthew Feickert}
\affil[14]{matthew.feickert@cern.ch, University of Illinois at Urbana-Champaign, USA, 0000-0003-4124-7862}
\author[15]{Clemens Lange}
\affil[15]{clemens.lange@cern.ch, CERN, Switzerland, 0000-0002-3632-3157}
\author[16]{Kilian Lieret}
\affil[16]{kilian.lieret@lmu.de, Ludwig Maximilian University of Munich, Germany, 0000-0003-2792-7511}
\author[17]{Rob Quick}
\affil[17]{rquick@iu.edu, Indiana University, USA, 0000-0002-0994-728X}
\author[18]{Arturo S\'{a}nchez Pineda}
\affil[18]{arturos@cern.ch, ICTP, INFN and University of Udine, Italy, 0000-0001-8241-7835}
\author[19]{Christopher Tunnell}
\affil[19]{tunnell@rice.edu, Rice University, Houston, USA, 0000-0001-8158-7795}


\begin{abstract}
New facilities of the 2020s, such as the High Luminosity Large Hadron Collider (HL-LHC), will be relevant through at least the 2030s.
This means that their software efforts and those that are used to analyze their data need to consider sustainability to enable their adaptability to new challenges, longevity, and efficiency, over at least this period.
This will help ensure that this software will be easier to develop and maintain, that it remains available in the future on new platforms, that it meets new needs, and that it is as reusable as possible.
This report discusses a virtual half-day workshop on ``Software Sustainability and High Energy Physics'' that aimed 1) to bring together experts from HEP as well as those from outside to share their experiences and practices, and 2) to articulate a vision that helps the Institute for Research and Innovation in Software for High Energy Physics (IRIS-HEP) to create a work plan to implement elements of software sustainability.
Software sustainability practices could lead to new collaborations, including elements of HEP software being directly used outside the field, and, as has happened more frequently in recent years, to HEP developers contributing to software developed outside the field rather than reinventing it.
A focus on and skills related to sustainable software will give HEP software developers an important skill that is essential to careers in the realm of software, inside or outside HEP.
The report closes with recommendations to improve software sustainability in HEP, aimed at the HEP community via IRIS-HEP and the HEP Software Foundation (HSF).
\end{abstract}

\begin{document}

\flushbottom
\maketitle
\thispagestyle{empty}


\section{Introduction}


New and being-developed facilities of the 2020s, such as the High Luminosity Large Hadron Collider (HL-LHC), will be relevant through at least the 2030s.
This means that their software efforts and those that are used to analyze their data need to consider sustainability to enable their adaptability to new challenges, longevity, and efficiency, over at least this period.
Considering sustainability in software development will help ensure that it will be easier to develop and maintain, that it remains available in the future on new platforms, that it meets new needs, and that it is as reusable as possible.
Software sustainability practices could lead to new collaborations, including elements of HEP software being directly used outside the field, and, as has happened more frequently in recent years, to HEP developers contributing to software developed outside the field rather than reinventing it.
Finally, a focus on and skills related to sustainable software will give HEP software developers an important skill that is essential to careers in the realm of software, inside or outside HEP.

To address this challenge, the first four authors of this paper organized a virtual half-day workshop on ``Software Sustainability and High Energy Physics''\footnote{\url{https://indico.cern.ch/event/930127/}}.
This workshop had two complementary goals:
\begin{enumerate}
\item To bring together experts from HEP, as well as those from outside, to share their experiences and practices, and
\item To articulate a vision that helps the Institute for Research and Innovation in Software for High Energy Physics (IRIS-HEP) in creating a work plan to implement elements of software sustainability.
\end{enumerate}
Eighty nine people registered for the workshop, though as a virtual workshop, attendees came and went at various times. We estimate that there were about 70-80 participants in the workshop at any one time.

This report discusses the workshop, and is organized similarly to the workshop, as shown in Table~\ref{tab:agenda}. The remainder of this section and the next two (\S\ref{sec:HEPexperiences} and \S\ref{sec:other_experiences}) summarize the talks given during the workshop, which were presented by the the authors of this report.
The final section (\S\ref{sec:discussion}) reports on the breakout groups that discussed future plans, and the plenary planning discussion that followed.

\begin{table}[ht]
\centering
\begin{tabular}{rll}
\toprule
09:00 & IRIS-HEP Blueprint process & Mark Neubauer \\ 
09:05 & Introduction to software sustainability & Daniel S. Katz \\ 
\addlinespace[1.3ex]
09:20 & Experiment experiences & Edward Moyse \\ 
09:35 & Experiment experiences II & Danilo Piparo \\ 
09:50 & HSF: HEP Software Foundation & Graeme A Stewart \\ 
10:05 & Community Software Successes \& Failures & Elizabeth Sexton-Kennedy \\ 
\addlinespace[1.3ex]
10:25 & Software Sustainability Institute (SSI) & Neil Chue Hong \\ 
10:35 & Astropy & Adrian Price-Whelan \\ 
10:45 & The Carpentries & Erin Becker \\ 
10:55 & HSF training & Samuel Meehan \\ 
11:05 & Research Software Engineers & Ian Cosden \\ 
11:15 & Software maintenance and capacity building in HEP & Ketevi Adikle Assamagan \\ 
\addlinespace[1.3ex]
11:25 & Discussion & breakout groups \\  
12:15 & Reports from breakouts & plenary discussion \\ 
12:30 & Planning next steps & plenary discussion \\ 
13:00 & Adjourn & \\
\bottomrule
\end{tabular}
\caption{Workshop Agenda (times are CDT) Presentations are available from the workshop website\textsuperscript{*}. \\
  \small\textsuperscript{*}\url{https://indico.cern.ch/event/930127/timetable/}
  \label{tab:agenda}}
\end{table}

\subsection{IRIS-HEP Blueprint process} 

The goal of the Institute for Research and Innovation in Software for High-Energy Physics (IRIS-HEP) is to address key computational and data science challenges of the HL-LHC and other HEP experiments in the 2020s. IRIS-HEP resulted from a 2-year community-wide effort involving 18 workshops and 8 position papers, most notably a Community White Paper~\citep{Albrecht2019} and a Strategic Plan~\citep{Elmer:2017rej}. The institute is an active center for software R\&D, functions as an intellectual hub for the larger community-wide software R\&D efforts, and aims to transform the operational services required to ensure the success of the HL-LHC scientific program. 

The IRIS-HEP Blueprint activity is designed to inform development and evolution of the IRIS-HEP strategic vision and build (or strengthen) partnerships among communities driven by innovation in software and computing. The blueprint process includes a series of workshops that bring together IRIS-HEP team members, key stakeholders, and domain experts from disciplines of importance to the Institute's mission. This blueprint meeting on the topic of \emph{sustainable software for HEP} is one of a series of workshops that have also included
\begin{itemize}
    \item Analysis Systems R\&D on Scalable Platforms (2019) \vspace{-8pt}
    \item Fast Machine Learning and Inference (2019) \vspace{-8pt}
    \item A Coordinated Ecosystem for HL-LHC Computing R\&D (2019) \vspace{-8pt}
    \item Software Training (2020)
\end{itemize}
The blueprint workshop discussions are captured and inform key outcomes which are summarized in a short report made publicly available, such as this report. 

\subsection{Introduction to software sustainability} 

The reason software sustainability is important is that software stops working eventually if is not actively maintained.

Generally, the structure of computational science software stacks~\citep{Hinsen_2019} is:

\begin{enumerate}
\item Project-specific software: software to do a computation using building blocks from the lower levels: scripts, workflows, computational notebooks, small special-purpose libraries and utilities  \vspace{-8pt}
\item Discipline-specific software: tools and libraries that implement disciplinary models and methods  \vspace{-8pt}
\item Scientific infrastructure: libraries and utilities used for research in many disciplines  \vspace{-8pt}
\item Non-scientific infrastructure: operating systems, compilers, and support code for I/O, user interfaces, etc.
\end{enumerate}
Software builds and depends on software in all layers below it. Any change in an underlying layer may cause the software to collapse.

Given this, we can define research software sustainability as the process of developing and maintaining software that continues to meet its purpose over time, which includes that the software adds new capabilities as needed by its users, responds to bugs and other problems that are discovered, and is ported to work with new versions of the underlying layers, including software as well as new hardware.

In order to sustain research software, we can
\begin{itemize}
    \item do things that reduce the amount of work needed, \vspace{-8pt}
    \item do things that increase the available resources, or \vspace{-8pt}
    \item do things that both reduce the amount of work needed and increase the available resources.
\end{itemize}

To reduce the amount of work needed, we can train developers, which involves finding or developing training material (see \S\ref{sec:hsf-training}). We can also use best practices, which involves finding or developing best practices.

There are a number of potential things we can do to increase the available resources.
We can create incentives that reward people who contribute to the software.
One specific example is to make the software citable, to make contributors authors, and to encourage users to cite the software, so that for developers in research institutions, they are rewarded through citations, which usually match the existing metrics on which they are hired and promoted.
We can also attempt to change career paths and associated metrics, by either adjusting existing career paths so that they reward software work, perhaps by adding new metrics, and by developing new career paths that focus on and reward software work, such as for Research Software Engineers (see \S\ref{sec:RSE}).
Though it is a long-term activity, we can try to increase the funding available for software work by first making the role of software in research clear to research funders, and then by clearly making the case for them to increase funding for new software, and to provide funding for software maintenance.
Finally, we can seek institutional resources for software that is considered sufficiently important to the institution, either  operationally or for its reputation, and we can demonstrate to our own institutions when other institutions do this.

To both reduce work and bring in new resources, we can encourage collaboration.
For example, using the work of others rather than reimplementing a function or package reduces what a software team (or its developers) needs to do themselves, even without assuming that the collaborators contribute to the software, which also may happen.
Similarly, if others use a team's software and contribute to maintaining it, the team has less they need to do.
To make this work, the software has to be designed from the start to be modular and reusable, and it must also be clearly documented and explained to potential users, even those in fields other than the developer’s.
And the team has to put effort into engaging and working with the potential user and contributor community.

Given that volunteers and collaborators are important to a number of sustainability paths, it's worth considered why volunteers or collaborators choose to put effort into a software project, and thinking about how we can engage them. 
In the context of community activities and organizing, \cite{Porcelli_2013} defines:

\begin{quote}
    Engagement = intrinsic motivation + extrinsic motivation + support - friction
\end{quote}

Where intrinsic motivation includes self-fulfillment, altruism, satisfaction, accomplishment, pleasure of sharing, curiosity, and making a real contribution to science; extrinsic motivation includes things like a job, rewards, recognition, influence, knowledge, relationships, and community membership; support means ease, relevance, timeliness, and value; and friction is technology, time, access, and knowledge.

Some examples of things we can do to increase engagement include the following.
Use GitHub (or GitLab) for development; this reduces friction by using a technology known to most developers.
Provide templates for issues and guidelines for good pull requests; this reduces friction by providing knowledge of how to work with our project, and increase support by easing the means of doing so.
Provide a code of conduct and a welcoming and encouraging environment; this increases extrinsic motivation by helping develop relationships and a sense of community.
Add contributors to a list of authors who are cited when the software is used; this increases both intrinsic motivation and extrinsic motivation through recognizing accomplishments.
Highlight examples of how the software is used; this increases intrinsic motivation by demonstrating the contribution to science.

\begin{figure}[ht]
\centering
\includegraphics[width=0.9\linewidth]{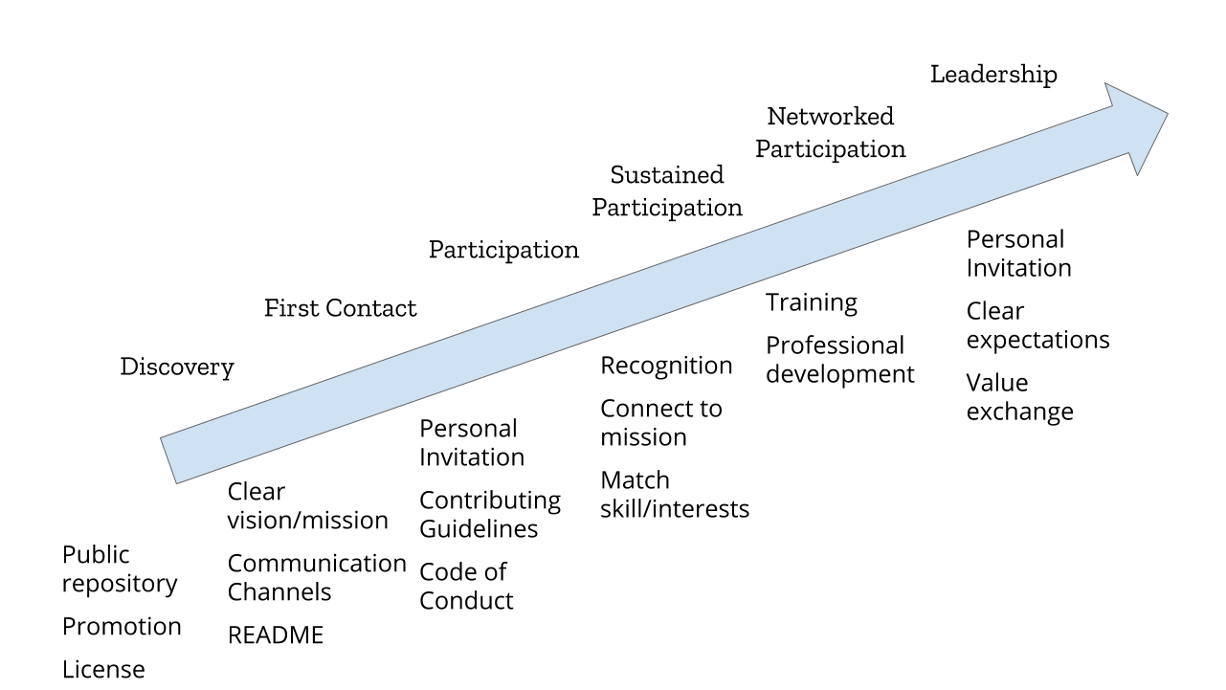}
\caption{How a project can encourage the potential contributor to move to from level to another \citep{Mayes_2020}.}
\label{fig:arrow}
\end{figure}

In addition to the general activities, we can also plan for a progression of types of engagements, as proposed by \cite{Mayes_2020} and as shown in Figure~\ref{fig:arrow}, with the goal of engaging a potential contributor in their first interaction with the software, and then moving their interaction to a higher and more significant level over time through intentional activities.

We should also remember that the challenge of software sustainability is not unique to HEP, and there are a number of other groups also working on this problem. In software sustainability generally, these include the
Software Sustainability Institute (SSI) (see \S\ref{sec:SSI}) and
US Research Software Sustainability Institute (URSSI) Conceptualization\footnote{\url{http://urssi.us/}}. 
The Research Software Alliance (ReSA)\footnote{\url{https://www.researchsoft.org}} is working to coordinate and align these organization with an interest in research software.
There are a number of different groups working on Research Software Engineering (RSE) (see \S\ref{sec:RSE}), including
the Society of Research Software Engineering\footnote{\url{https://society-rse.org/}}, 
the US-RSE Association\footnote{\url{http://us-rse.org/}},
and groups in other countries/regions (DE, NL, Nordic, BE, AUS/NZ)\footnote{\url{https://researchsoftware.org/assoc.html}}.
Finally, working on training and education is The Carpentries (see \S\ref{sec:Carpentries}).

\section{HEP experiences}\label{sec:HEPexperiences}

The first session of contributions focused on general software experiences from the HEP community, including: ATLAS and CMS, the two largest experiment collaborations; the HEP Software Foundation (HSF), an organization that facilitates cooperation and common efforts in High Energy Physics software and computing internationally; and some lessons about previous software experiences.

\subsection{ATLAS experiences}\label{sec:ATLASexperiences} 

ATLAS is a large collaboration of around 2,900 scientific authors and around 1,200 students spread around the world. This situation is far from unique in HEP, but it does present some challenges, such as the complexity of managing a distributed team of developers (especially given the lack of traditional carrots and sticks found in industry).

ATLAS initially used many home-grown tools (or at least, tools that were exclusive to HEP). Examples include CMT\footnote{\url{http://www.cmtsite.net}}, TagCollector~\citep{Albrand}, and CLHEP\footnote{\url{http://proj-clhep.web.cern.ch}}. The advantage was that, at least in principle, we get tools and libraries that are perfectly tailored to our use case. The disadvantage is that we have an ongoing maintenance burden for software outside of our core tasks, and also that dedicated training is needed for non-standard tools.

In recent years, ATLAS has moved to more modern, industry-standard options, such as git + CMake, Eigen, etc. This means developers learn transferable skills, can benefit from excellent online tutorials, and in many cases, the community benefits from better written software. Reducing the amount of extraneous work we need to do is a huge step towards being able to focus on the sustainability of our core software.

Athena~\citep{athena} is ATLAS's event processing framework, and consists of $>$1 million lines of Python and $>$4 million lines of C++.
We open-sourced Athena at the end of run-2 (late 2018), which made it easier to collaborate with industry, and others in the field, once again reducing the burden on ATLAS's developers. Other examples of open-sourced software originating from ATLAS include Rucio\footnote{\url{https://rucio.cern.ch}}, GeoModel\footnote{\url{http://geomodel.web.cern.ch/geomodel}}, ACTS~\citep{acts} and Phoenix~\citep{phoenix}. 

Within ATLAS, we make extensive use of social coding features, such as merge (pull) request reviews and continuous integration. We have two levels of shifters, working in morning and afternoon shifts, and they review approximately 40 merge requests per day. These reviews are vitally important to improving the quality (and sustainability) of our software.

We also run various checks on our software every night. We build about 20 different branches of Athena, and on each of these we run:
\begin{itemize}
    \item Unit tests,  \vspace{-8pt}
    \item Longer local tests,  \vspace{-8pt}
    \item Grid-based large statistics tests,  \vspace{-8pt}
    \item Checks on technical performance (CPU and memory).
\end{itemize}
The primary purpose of the non-unit tests is to catch more subtle bugs and regressions. And we also run even larger (around 1 million events) validation campaigns to measure physics performance (and find very, very rare bugs.)
 
Merge request reviews only examine code that is changing, but we periodically also use various static analysis tools, such as cppcheck, lizard, and Coverity, to examine the entire codebase.
 
For documentation, we use TWiki, and also have some dedicated expert-maintained documentation, which is, unlike the content on Twiki, public and visible to search engines. 

Writing modern, maintainable code is very important. To this end, we run various training campaigns: every new ATLAS person is strongly encouraged to go to a week-long induction, which includes an introduction to software development. We also run more infrequent training for core developers, and on particular software topics.

One key problem we experience with maintaining our software is retaining our experts. Many leave because they cannot find a job in the field. We have made efforts to combat this, with software grants and trying to encourage institutional commitment to maintenance tasks, but ultimately the solution is that funding agencies prioritise hiring software experts.

\subsection{CMS experience}\label{sec:CMSexperiences} 

Computing and offline software is central to CMS data collection and processing operations, Monte Carlo production, and physics analysis, and is orchestrated over hundreds of thousands of cores. Key elements of software sustainability are somewhat built into its very design. 

The CMS software has always been open source and has been hosted on GitHub for almost a decade\footnote{\url{https://github.com/cms-sw/cmssw}}. The CMS software stack also includes several third-party packages that are also open source, such as ROOT and Geant 4 (Monte Carlo event generators). 

While CMS is attentive to software support and sustainability in the Run 4--5 timescale (2027--2030 and 2032--2034), the challenges posed by HL-LHC invite CMS to explore solutions that might be disruptive. It is desirable to model future evolution based on as much common tools as possible across HEP and even outside it.

Prerequisites for CMS event processing and computing software are cost effective computing architectures, and being able to rebuild on different platforms, adapt to CMS needs, and rely on existing community of developers and users. While the exact details of the microarchitectures on which the code will run in 10 years from now are unknown, a prerequisite for being forward compatible is to avoid any platform-specific code or other intrinsic elements. Continuous integration on different platforms ensures an early detection of bugs in the code and improves its quality, numerical stability and memory usage patterns. For aspects related to toolkits for detector simulation and description, and data and workload management infrastructures, CMS values the adoption of standard tools where possible. Examples are Rucio for data management, CRIC (Computing Resource Information Catalogue) as an information system, and DD4hep as a geometry description toolkit.

CMS is committed to addressing the challenge of disk storage needed to support data analysis. The experiment has been able to drastically shrink the size of its analysis datasets: the size of an event in the NanoAOD format\footnote{\url{https://www.epj-conferences.org/articles/epjconf/pdf/2019/19/epjconf_chep2018_06021.pdf}} is about 1--2 kB. This format is now produced for all the CMS data and Monte Carlo samples, and CMS plans to increase its utilization in the next years.

CMS believes that these approaches will empower the long-term sustainability of its software.

\subsection{HEP Software Foundation}\label{sec:HSF} 

As many of the problems faced by the scientific community regarding software are
quite general it is natural to take a whole community approach to solving them.
To this end the HEP Software Foundation\footnote{\url{https://hepsoftwarefoundation.org/}} was founded in 2015 to
encourage a cooperative attitude towards improving software in the High Energy
Physics domain.

One of the first tasks undertaken was to survey the field to establish a roadmap
for the next decade's work~\citep{Albrecht2019}, including the specific issues
surrounding software development and tooling~\citep{couturier2017hep}, which inform
many of the points made here.

Firstly, we strongly encourage developers to consider re-using existing software
packages and projects, contributing any missing features, rather than embarking
on developments that duplicate existing tools. Efforts to establish catalogues
of software to help with searches for existing software have often fall foul of
not finding dedicated or community effort to sustain them, although the HSF
is working with the ESCAPE project\footnote{\url{https://projectescape.eu/}} to attempt this again in a more
sustainable way.

When starting new software projects it is far easier to begin well than to have
to correct defects post facto. The HSF provides a set of best practice
recommendations~\citep{hegner_benedikt_2020_3965581} and even has a
template\footnote{\url{https://github.com/HSF/tools}} that can be used to setup a project with many of
the right pieces done correctly (e.g., hooks for tests, documentation, license,
etc.). For projects written in C++, CMake has become the defacto standard for
the build system and there is an excellent guide to modern CMake
practice~\citep{ModernCMake}. Similarly for Python projects the Scikit-HEP project
provides a useful guide for developers\footnote{\url{https://scikit-hep.org/developer}}. One thing that
should be established from the outset is the copyright holder for the code and
the licensing terms. For copyright, HEP projects can usually usefully assign
this to the host laboratory, e.g., CERN, whose scientific mission is to ensure
dissemination of knowledge. This is achieved by licensing the software as
open-source. There are many licenses that are considered open-source and the HSF
has a guide to help developers understand the
options~\citep{jouvin_m_2016_1469636} options. More liberal licenses (non-GPL) are
usually preferred as these provide the most flexibility for users of the
software to combine with other software and to license their own software on the
terms that they would like.

The software used in HEP experiments these days consists of many components,
from generic libraries to very domain specific components, and these are built
into a software stack. Ensuring integration with other software, and a clear
understanding of dependencies, will make the life of software librarians easier
and help adoption of the software. Spack~\citep{7832814} is one popular build tool
and orchestrator that is popular in the scientific community and is being used
to provide turnkey stack solution as part of the CERN Experimental Physics Department's R\&D and HSF's Key4hep
projects\footnote{\url{https://test-ep-rnd.web.cern.ch/topic/software/turnkey-software-stack}}. For software that forms part of the smaller stack
used by analysers Conda is a popular alternative\footnote{\url{https://conda.io/}}. Solving this
dependency and orchestration problem makes it easier for end-to-end domain
solutions to take the form of a toolkit, consisting of smaller, more focused
pieces that are easier to attract contributions to, but also easier to replace
in the overall scheme if that proves to be necessary.

All sustainable software evolves over time and there is one particular challenge
that scientific software in general, and HEP software in particular, faces
today. This is the change of hardware away from CPUs and towards accelerators,
such as GPUs. Such devices rely on less sophisticated programming models than
C++ and there is a large, and evolving, variety of hardware available. Making
software sustainable and evolving it appropriately in these circumstances
is difficult. The community is now putting significant efforts into finding the best 
sustainability APIs, based on the needs of HEP\footnote{\url{https://indico.cern.ch/event/908146/contributions/3826737/}}, but this
remains an open point to keep an active watch on.

Finally, it should be emphasised that it is a learned skill to write the kind of
good, well structured, software that forms part of a sustainable ecosystem. To
that end good training must be provided and making such training available as a
curriculum for people entering the field, particularly at more junior levels,
is vital. The HSF Training Working Group has recently been highly active in
developing such training materials and running such courses, as discussed more in
\S\ref{sec:hsf-training}.

\subsection{Community Software Successes \& Failures} 

The field of high energy physics has a long history of community software developed for the greater good of supporting science without commercial pressures, and allowing scientist to focus on software that processes their unique scientific instruments instead of implementing, yet again, common mathematical functions.  CERN's CERNLib is probably the oldest example of this, written over 40 years ago in the late 1960s and 1970s in FORTRAN66 and 77 \citep{web-grid}.  It only faded away in the 2010s\footnote{Some may argue it is not dead yet, however the most recent port to scientific Linux 6 in 2012 was its last.}.   If Tim Burners-Lee did not have this greater good attitude in mind, the World Wide Web would not have developed as an open and free system.

Because of this long history, it is possible to examine many software projects both successful and the unsuccessful.  For the unsuccessful one can ask, ``Why did it die?'', and for the successful, one can ask, ``What is it about this project that has sustained it over the years? Are there valuable lessons to be learned from its story?''  Projects that have lasted more then 10 years are the most interesting in this respect.  In a survey of such projects, the software developers and the institutions that support their careers are the most significant factor in guaranteeing the longevity of a project.  Another significant factor is open software; it is a requirement of sustainability.  Many operating systems have come and gone, some even superior like VAX/VMS, but Linux is still with us because it is open and freely available.  That openness allows multiple institutions and companies to sustain it through the decades.  An example of multi-institutional software development in HEP is INSPIRE.  SLAC, IN2P3, IHEP, Fermilab, DESY, and CERN all contribute to the software and sustain the service as it is so intimately tied to the mission of these institutions.  Other multi-institutional products include dCache, RUCIO, and the children of CERNLib: Geant4 and ROOT.

It is possible for a single institution to sustain a software product if it is closely tied to the mission of that institution and it has thousands of users.  Examples of this in HEP are INDICO, Enstore, Frontier, FTS, and HTCondor.  Most of these are supported at HEP laboratories that can create careers  and succession plans for the developers that sustain and modernize the product.  The very interesting exception is HTCondor.  This is an example of a product sustained by a single US funding agency, at a University by the commitment and force of personality of its PI.  It is a wildly successful batch scheduling system used all over the globe.  The common element for these single institution products is the people involved.  Counter examples of products that are fading away are PhedEx, HPSS, and gridFTP.  There are no clear owners of these software packages.  PhedEx was under-supported and the CMS experiment decided to move to the community supported RUCIO tool.  HPSS is indifferently maintained by IBM and the HEP community believes gridFTP will die out in this community because the Globus tool kit has been moved to closed source. 

The above gives two models of supporting software over many decades.  However both demonstrate that it is the people involved that determine longevity success or failure.   

\section{Software, Training, Careers} \label{sec:other_experiences}

The second session of contributions focused on more general software experiences, including: the Software Sustainability Institute, a UK organization that works to improve research software; Astropy, a community that develops software for astronomy; The Carpentries, a community that develops and delivers training material, including for research software; the HSF and IRIS-HEP training activities that use The Carpentries model with content specialized for HEP; research software engineering, a specialized role that sits at the intersection of research and software engineering; and 
the Snowmass 2021 activity as a means to build community consensus around some of these ideas and potential solutions.

\subsection{Software Sustainability Institute (SSI)}\label{sec:SSI} 

The Software Sustainability Institute (SSI) \citep{Crouch_2013} was established in 2010, with support from the UK research councils, to improve the quality of software used by researchers. Over the last ten years, it has provided a wide range of resources -- including expertise, services, tools, events, policy, guidance, data and opportunities -- creating a national facility for cultivating better, more sustainable, research software to enable world-class research.

The work of the SSI is split into five areas: software consultancy, training, community building, policy, and outreach. When the SSI first started, the focus was on consultancy. However, to increase the impact of its work, it was clear that other activities were required that scaled up the number of people who could benefit. By delivering training and establishing communities of practice, who could then takeover the work of improving research software practice themselves, the SSI was able to reach a much wider audience and gain from bringing together the community to identify and address key challenges, such as recognition for software development and maintenance. By collecting evidence on the research community's use of software, it was possible to argue for changes in policy. Each part of the SSI informs the work of the rest.

Part of this strategy is helping focus individual efforts, knitting them together, and amplifying them. We helped universities in the UK deliver Carpentries training workshops, coordinating requests for instructors and supporting local hosts. This has gone from 7 workshops in 2012 to 57 in the last 12 months, with the UK contributing to the global Carpentries community by developing and maintaining lessons. Our Fellowship program \citep{Sufi_2018} has helped bring together a cohort of over 150 individuals championing software practice in their own domains. This has enabled them to work together to deliver domain-specific initiatives or tackle challenges such as helping make research open and reproducible. The role of the Research Software Engineer was born\footnote{\url{https://www.software.ac.uk/blog/2016-08-17-not-so-brief-history-research-software-engineers-0}} from a discussion session at the SSI's Collaborations Workshop in 2012 on how to achieve recognition for working on research software and, with coordination and backbone support from the SSI, has become a worldwide movement that has led to the formation of new job titles, career paths and a professional society\footnote{\url{https://society-rse.org/}}.

However, to achieve cultural change, it is also necessary to change systems and policy. The SSI has supported this through the collection and publishing of data, including the 2014 UK Research Software Survey \citep{2014_Research_Software_Survey} and ongoing international RSE surveys \citep{2018_International_RSE_Survey}. This feeds into the campaigning work the SSI does to gain recognition for the fundamental and underpinning role software plays in research, and the development of policies and guidance - in collaboration with other organizations - to ensure that software is reliable, reproducible and reusable.

Research software sustainability, computational and data skills, and career paths for research software professionals are intrinsically linked. The Software Sustainability Institute was one of the first organizations working to improve practice in this space, but it can't be the last. To achieve a lasting impact, in domains such as High Energy Physics, there must be support for the formation of communities of practice to ensure the sharing, learning and growing of practice and process around software sustainability.  

\subsection{Astropy} 

Astropy\footnote{\url{https://www.astropy.org/}} started as a community-developed Python library aimed at helping astronomers perform their research \citep{astropy:2013}.
This library (i.e., the installable Python package \texttt{astropy}) is both open source and open development, and contains core functionality that all or most astronomers use on a regular basis.
Astropy has since grown into the broader, community-oriented ``Astropy Project,'' which is both an ecosystem of software packages aimed at astronomers and physicists (including the Astropy core library and other, domain-specific packages) \emph{and} the community that supports and uses this code (i.e., the developers, maintainers, users, educators, researchers, etc.).
The core Astropy library now serves as a base package and exemplar of the Project's values (community oriented, open source, open development) upon which more focused packages are built \citep{astropy:2018}.
In the $\sim$8 years since its inception, Astropy and the affiliated ecosystem of packages have become a fundamental part of the astronomical research toolkit (for example, the Astropy core library summary papers currently receive about 100 citations per month, but the rate is still increasing).

The Astropy Project contains an ecosystem of packages, but this is an ecosystem within an ecosystem: The broader set of more general Python packages that enable the ``Python scientific software stack.''
Within this context, the Astropy ecosystem is built on the Astropy core library, which is built on and parallel to the scientific analysis and visualization ecosystem (e.g., scipy, matplotlib, pandas), which is built on the core numerical and functional ecosystem (e.g., numpy, Jupyter, Cython).
In reality, these ecosystems are not as stacked as they are inter-connected at the software level, but it is important to acknowledge that the communities  oriented around these ecosystems may not be as connected.
From the perspective of software sustainability, this can become a problem, both because it can create a disconnect between large user groups and developers, but also because it inhibits spreading maintenance and support work over a larger pool of contributors.

A key goal of the Astropy Project is to empower scientists and scientific software developers to write their own software within the Astropy ecosystem, but to also encourage communication and contribution back upstream through the whole software stack.
One of the ways we enable this is by providing a package template\footnote{\url{http://github.com/astropy/package-template}} to help scientists package their code, which implicitly encourages them to adopt the open source values and the software standards (for testing, documentation, and packaging) of the Astropy Project.
However, there are many other key aspects of software sustainability that go beyond sustaining the software stack and Python ecosystem. In particular, within Astropy, we have focused on:
\begin{itemize}
    \item maintaining code and adapting it for continued utility,  \vspace{-8pt}
    \item supporting contributors and training new ones, and  \vspace{-8pt}
    \item educating and growing the user base.
\end{itemize}

For the code, we prioritize developing and maintaining the infrastructure behind the code, testing the code with continuous integration services, providing regular releases, and developing new features driven by community input.
As an open source and open development project, most of our work is done on GitHub, tested with continuous integration services (CircleCI, TravisCI), and our documentation is built automatically with Sphinx on ReadtheDocs.

For contributors, we have an explicit code of conduct\footnote{\url{https://www.astropy.org/code_of_conduct.html}}, we are now beginning to fund core maintainers, and we have recently established a new, clearer governance structure for the project\footnote{\url{https://github.com/astropy/astropy-APEs/pull/61}}.
Our code of conduct has been critical for establishing clear expectations for our contributors, maintainers, and users, both about how to behave and in defining responsibilities.
Lastly, while historically much of the development work within Astropy has been done in volunteer time, we now have funding from the Moore Foundation for Sustaining and Growing the project (which we are using to pay developers, to establish a better contributor pipeline, and to formalize our governance structure).

For supporting and training users, we run workshops as a part of career development tracks at conferences (e.g., the American Astronomical Society meetings), and provide many points of access between users and maintainers (mailing lists, Slack group, GitHub issues).
We have also found that the open development model that we use has naturally created many entry points into the user-to-maintainer pipeline, allowing people with different experiences and training to join the Project in different capacities.
While we have many success stories with empowering users to become ``bug reporters'' or educators, we are still in the process of figuring out how to foster the next steps in the pipeline (empowering ``power-users'' to contribute and take on leadership roles in the project).

A large component of why the Astropy Project has been sustainable thus far is that we exist within an established ecosystem that is maintained by, and with, other communities (i.e., the scientific Python stack).
However, the Astropy Project is now consciously fostering long-term sustainability by focusing on building our community, and empowering that community to lead and drive itself.

\subsection{The Carpentries} \label{sec:Carpentries}

The Carpentries is a community-led project\footnote{The Carpentries is a fiscally-sponsored project of Community Initiatives, a 501(c)3 non-profit based in California, USA.} that focuses on developing and providing training in the core computational and data skills for efficient, shareable, and reproducible research. Short, intensive, hands-on Carpentries workshops give researchers the opportunity to engage in deliberate practice as they learn these skills, starting with strong foundational skills and receiving feedback as they learn. Carpentries workshops are designed for people with little to no prior computational experience and teach not only an introduction to programming, but also the perspectives and skills for developing software in an applied context. This model has been shown to be effective, with the vast majority (more than 90\%), of learners saying that participating in the workshop was worth their time and led to improvements in their data management and data analysis skills. Since 2012, The Carpentries has trained over 57,000 learners in 61 countries and trained more than 2,400 volunteer instructors to deliver over 30 collaboratively-developed, openly-licensed lessons. In addition, The Carpentries supports its community in developing, sharing, and collaborating on lessons beyond its core curricula, with its Incubator and CarpentriesLab projects\footnote{\url{https://carpentries.org/community-lessons/}}. As of September 2020, these projects contain more than 30 contributed lessons on topics ranging from containerization to fMRI image analysis.

As a volunteer-led project, the success of The Carpentries is predicated on its ability to sustainability engage and support an extensive network of volunteers with a wide variety of skill sets, technical expertise levels, and domain backgrounds. Its sustainability model rests on six core principles: 

\begin{enumerate}
    \item Gather people with shared values - The Carpentries community grew organically around an implied set of values, which The Carpentries has since formally identified through a community-driven process, and articulated\footnote{\url{https://carpentries.org/values/}}. The Carpentries focuses on creating a welcoming environment for individuals who already share these values.  
    \item Provide multiple avenues to contribution - The Carpentries recognizes that people have different capabilities and access to different resources. It strives to provide a wide range of options for people to get involved - to maximize community energy and accessibility.
    \item Value all contributions - The strength of The Carpentries lessons depend on continuous improvement from its community. The Carpentries welcomes all contributions, from typo corrections to creating entire new curricula, and each of the 2400 instructors have contributed to the lessons.
    \item Provide growth opportunities - The Carpentries works with its volunteers to provide valuable skills that they can use for their current or future jobs. Depending on individual interests and career trajectories, this ranges from pedagogical training and teaching experience to experience with specific technologies and community development.
    \item Reward volunteerism - The Carpentries recognizes its volunteers through multiple avenues, including certification, authorship credit, community service awards\footnote{\url{https://carpentries.org/awards/}}, and visibility on our website\footnote{\url{https://carpentries.org/instructors/}}. The Carpentries works with volunteers' employers to recognize their volunteer work through memoranda of understanding. 
    \item Focus on resource creation - Lastly, The Carpentries focuses on producing high-quality documentation and other resources for every aspect of its work. By externalizing the expertise of its community members, The Carpentries lowers barriers to involvement and increase the ability of its community to carry out their mission.
\end{enumerate}

These six aspects of its sustainability model have enabled The Carpentries to grow from a handful of instructors teaching a few workshops a year to a global community of thousands, collaboratively developing and delivering nearly 500 high-quality training events for 12,000 learners per year.

\subsection{HSF training}\label{sec:hsf-training} 

Having sustainable software requires that the research community be educated such that they can fully harness the capabilities of software in the present and that this expertise persist as senior generations retire and are replaced by younger generations of scientists.  The HEP Software Foundation appreciates this and has formalized its role by creating a group dedicated to training\footnote{\url{https://hepsoftwarefoundation.org/workinggroups/training.html}} with the mission \textit{to help the research community to provide training in the computing skills needed for researchers to produce high quality and sustainable software}.  This group develops training materials and executes workshops with a pedagogical approach similar to that of the Software Carpentries\footnote{\url{https://software-carpentry.org/}} rooted in five principles: 1) hands-on, 2) student-centric, 3) experiment agnostic, 4) re-useable, and 5) open and accessible.  

These lessons are cast in a style inherited from The Carpentries and housed on an open source HSF-training GitHub space\footnote{\url{https://github.com/hsf-training}} such that any individual can contribute. A comprehensive overview over all planned and completed training modules is given in the \emph{HSF Training Curriculum}\footnote{\url{https://hepsoftwarefoundation.org/training/curriculum.html}}. As lessons become more mature, video playlists are developed to facilitate a broader reach and virtual instruction.  These videos are housed on the HSF YouTube channel\footnote{\url{https://www.youtube.com/watch?v=Q-vuR4PJh2c}}.  The developed content aims to be complementary to that already available within the curriculum of The Carpentries and covers topics ranging from continuous integration to machine learning and code documentation.

Training events are held in two formats: 1) in-person and 2) virtual.  Each training event requires the participation of a \textit{Facilitator} to be the primary organizer, one or more \textit{Instructors} to develop the event's material and drive the event itself, and a cohort of \textit{Mentors} to serve as teaching assistants to provide individual instruction by ensuring that the ratio of participants to educators does not exceed five. In-person events are held for audiences of up to 50 participants and the training materials are presented and worked through in real time by the instructor with support from the mentors.  Virtual events are organized using the ``flipped classroom''\footnote{\url{https://en.wikipedia.org/wiki/Flipped_classroom}} paradigm in which individuals work through self-guided materials in the form of training video playlists that complement the lesson web page at their own pace, with support from educators on a virtual chat platform (e.g., Slack/Mattermost) and then gather in small groups with mentors after a period of a few days to debug any confusion or explore more advanced topics.  In both training formats, it is essential that there be time for both the core training materials as well as the ability to explore advanced topics and application/experiment specific issues. The level of learning is gauged by self-reported estimates by participants and in both formats, it is generally reported to be successful in achieving the learning goals of the training event. An in-depth guide to organizing training events in both formats is available on the HSF training website\footnote{\url{https://hepsoftwarefoundation.org/training/howto-event.html}}.

In the future, we would like to continue to develop the core content of our curriculum (e.g., ``C++ for HEP'' training), collaborate more closely with The Carpentries in the context of the Incubator\footnote{\url{https://carpentries.org/involved-lessons/}} to make the lesson content more available to a wider audience for both consumption and development, and understand what aspects of training in computing factorizes into and can be implemented through augmentation of university curricula.  Finally, we also feel that a concrete commitment through the creation of career paths and other incentives with enhanced focus on training will facilitate more effective future progress since at the moment all aspects of the HSF training group and their community of educators is performed on a voluntary basis by individuals who recognize its importance to our long-term success as a field.

\subsection{Research Software Engineers}\label{sec:RSE} 

The breadth and sophistication of software development skills required to build and maintain research software projects are increasing at an unprecedented pace.
Researchers thrust into the role of developing software, with little or no software development experience or training, often employ ad-hoc, potentially detrimental development methods.
It has become clear in recent years that the level of effort and required skills to keep pace with computer and programming tools is not in the repertoire of the average researcher~\citep{Merali-sci-computing}.
When novices develop software or when researchers are more focused on research publications than on producing quality software, problems can arise that limit its usability, sustainability, and even accuracy ~\citep{science-retraction,python-os-bug}.
It is not uncommon for software tools and research code to become unusable after a project ends or the primary developer leaves.

One solution to the problems facing research software is the emergence of the Research Software Engineer (RSE).
Coined in 2012, the term Research Software Engineer has been broadly used as an inclusive title to describe anyone who understands and cares about both good software and good research\footnote{\url{https://society-rse.org/}}
\citep{Baxter}.
More specifically, an RSE is someone who views the development of research software as the primary output of their work efforts.
This distinguishes RSEs from domain researchers who view research publications as the primary focus of their work.

By combining an intimate knowledge of research with the skills of a professional software engineer, RSEs have the ability to transform traditional computational research by directly compensating for a domain researchers’ lack of software development expertise.
An experienced RSE has the tools and knowledge to allow them to work collaboratively with domain researchers in a manner that ensures the quality, performance, reliability, and sustainability of the software.

In addition to the software development and technical expertise they bring to a project, RSEs serve two other key roles in the research software ecosystem.
First, RSEs serve as leaders and mentors to novice software developers, including undergraduates, graduate students, and postdoctoral researchers.
By providing mentorship on research software projects, RSEs serve to elevate domain scientist’s development through exposure to professional best practices.
Second, RSEs are increasingly recruited to design and deliver software training programs to students and researchers.
These training programs are uniquely beneficial as RSEs understand the technology, audience, and requirements of research software. 
  
Often the most common entrance into an RSE career begins with an early-career researcher who shows interest in software development, or has it thrust upon them  \citep{uk-rse-nation}.
Recently, however, national and international RSE organizations have formed to support this important work and provide a community for people in RSE roles, advocate for the role, and support the formation and formalization of an RSE career path\footnote{\url{https://researchsoftware.org/assoc.html}}.

\subsection{Software maintenance and capacity building in HEP} 

The Particle Physics Community Planning Exercise, Snowmass 2021\footnote{\url{https://snowmass21.org}}, organized by the Division of Particles and Fields of the American Physical Society, is a scientific study that provides an opportunity for the particle physics community to come together to identify and document a scientific vision for the future of particle physics in the U.S. and its international partners. Snowmass includes ten Frontiers (focus areas). Two of these, the Community Engagement and Computational Frontiers, aim to discuss issues related to software sustainability, and to suggest recommendations for improvement.
 
These issues include software documentation as a key part of software maintenance. Well-documented software provides a good example for the people that inherit such software. This decreases the chances that they will rewrite it, which often happens if the software is not understood or liked, and such rewriting is a waste of time and effort. Good documentation also serve as a excellent educational platform when the software is passed on to the younger generation. Another key issue is increasing diversity and inclusion in the computing and software fields.

\section{Discussion}\label{sec:discussion}

After the talks, all remaining participants (about 25) were assigned to one of five breakout groups and asked to talk about the workshop goals, and to propose three specific actions that the HEP community could take to make software more sustainable.  At least two were intended to be things that could be accomplished within two years, or at least that would make significant progress that would lead to a measurable difference within two years.

\subsection{Ideas}

Group 1 focused on developing better software, including how to help developers ensure that their software worked on multiple platforms through continuous integration, perhaps including a collaborative effort to build a continuous integration platform for HEP. They also discussed how to improve documentation, including possible incentives such as awards and how documentation fits with citation. Finally, they discussed career paths for those who work in computing and software.

Group 2 started by discussing career paths, comparing academia, national laboratories, and industry. They also discussed how emphasizing software work affected these options, particularly as compared to emphasizing hardware work, which seems in part to be related to software sometimes not being seen as a core part of an experiment. Different HEP experiments also seem to have different policies, cultures, and practices with respect to the value of software and its developers and maintainers.
When talking about actions, they felt that it was important for software work to be rewarded through publication credit, and that the national laboratories are currently doing a good job with this. To go further, they suggested that all software created as part of experiments must be written up in a software paper and that this must be cited by experiments.
They also suggested that it was important to recognize that longer term positions are essential for software to be sustained, as otherwise knowledge is lost too quickly. To impact funding agencies, they suggest having HEP RSEs on the funding panels and on funder advisory committees. To impact universities and laboratories, they suggest having HEP RSEs on hiring committees. And to impact experiments, the suggest a ``service pledge'', where an institution needs to have software professionals and contribute their effort in order to have continued access to the collaboration (access to collision data, etc.)

Group 3 started in the present by discussing how COVID-19 had changed research, focusing on the rise of virtual meetings.
While on one hand, virtual meetings are more sustainable and scalable, lower cost and easier to participate in, they also don't involve the same commitment from attendees as when they travel to an event.
It's much easier to register for an event, particularly if there is no registration cost, that to actual commit the time for the event.
For example, the recent PyHEP conference had about 1000 people who registered, with actual attendance between 40 and 400 at various times.  Furthermore, the financial burden of participation in a given event is considerably lower as there are no costs associated with travel and housing and generally the registration fee is considerably lower or non-existent.
The group asked, given that current budgets include travel costs that are not being spent for travel, how could these budgeted funds be reused?
They examined ideas to incentivize participation by more people using money, particularly in carpentries-style training events.  One form this could take is to 
pay some participants as a means to get their undivided attention, which could be done selectively and could be used as a way to get participants to commit to being a training/mentor later on.
Similarly, mentors and instructors could be paid for their time and effort to help ensure a standard of quality that does not \textit{only} rely on ``the good will'' of those teaching.
This concept of paying for instruction can be extended to the development of training content, thereby allowing for a core HEP software curriculum to be developed more rapidly and reward, in a \textit{concrete} way, those who commit time and energy to doing so.
Another possible method to ensure commitment of attendees is to provide certificates of completion requests.

Group 4 discussed three separate topics.
First, the fact that C++ skills are needed for those that will do research, development, maintenance, and operations for software needs to be emphasized, particularly as there are performance-critical parts of data processing and analysis, where C++ delivers much faster code than Python.  Thus, C++ training needs to be brought to the same level of maturity and success as Python training, for the fraction of the software community who will need it. The FNAL C++ course is an example of a successful course, but it is not immediately scalable. Last year (2019), 50 people attended, and 30 completed the course, with positive feedback and a high level of engagement.
Second, given the increased use and sharing of containers, the HEP community needs to develop best practices (and possibly tools, or adopt industry solutions) for their use in sustainable software development.  
Third, there is a need to develop a plan to provide better incentives and mentoring for young scientists to engage in software and computing. Specifically, this need is for an appropriately balanced mixture of RSEs (professionals) and scientists working together to create a sustainable situation with both the software and pipeline of talent that works on the software. However, the appropriate incentives and professional development aspects are unclear.

Group 5 discussed a number of distinct topics.
One was the fact that application containers are not important, and one possible path to facilitate sustainability is to use Docker locally, with automatic conversion to facility-dependent HPC container technology, which requires an adoption of standards, such as the Linux flavor and version that is maintained inside the container.
A second idea was to use preexisting open software tools to generate project templates for new projects, including testing, documentation, and other software quality control procedures.
A third idea was to hold a workshop to identify places for common software, such as for data management, tracking software, user interfaces, and software delivery. This would include a systematic review of current codebases.
A fourth idea was to identify opportunities for services needed by HEP that experiments or other groups are interested in would co-develop, to eventually reduce costs. 
A fifth idea was to collaborate with organizations trying to professionalize software development and cyberinfrastructure in academic environments. This is currently being explored in some HPC centers and with infrastructure providers such as those providing gateway services. This would provide a career path for those researchers that take code development and cyberinfrastructure as their core role. 

\subsection{Potential actions}

After each group proposed its three actions, the participants collectively voted on them, with each participant asked to vote for three items.  Table~\ref{tab:actions} contains the actions and the votes.

\begin{table}[htb]
\centering
{
    \renewcommand{\arraystretch}{1.3}
    \begin{tabular}{>{\centering\arraybackslash}p{0.7 cm}p{11 cm}>{\centering\arraybackslash}p{1.4 cm}}
    \toprule
    \# votes & \multicolumn{1}{c}{Proposed action} & Proposing group \\
    \midrule
    10 &  Repurpose nominal funding from in-person training to pay content developers & 3 \\ 
    8 & Promote recognition and financial support/career paths: Have HEP RSEs on the funding panels / funder advisory committees / hiring committees & 2 \\ 
    4 & Incentives and professional development, some certification? RSEs training incoming HEP students. Mentoring & 4 \\ 
    4 & Organize a workshop to explore and define common software and services, including libraries, data and computational services, and gateways (VREs) and analysis services & 5 \\ 
    3 & Figure out how to develop career paths for the people who work in this & 1 \\ 
    3 & C++ training for HEP (that is sustainable and scalable). Modern C++. This is now a speciality. Make this equal with detector development specialism. & 4 \\ 
    2 &  Repurpose nominal funding from in-person training to pay participants & 3 \\ 
    2 & Incentivize training participant to instructor transitions & 3 \\ 
    2 & Use of containers (to support analysis; well developed in ATLAS). Sustainable and integrated with repo + CI. Develop best practices & 4 \\ 
    2 &  Define 2 or 3 common development environments as community accepted standards, then make these available via application container technology (e.g. Docker), and include project templates for testing, documentation and more. Proposed name (tongue in cheek): the HEP software development kit (HPSDK).  & 5 \\ 
    0 & Expand the use CI or other testing, to as many platforms as feasible \& reasonable & 1 \\ 
    0 & Figure out how to reward documentation & 1 \\ 
    0 & Create a policy at experiment/community level that all software created as part of experiments must be written up in a software paper (something that gets a DOI) and that this must be cited by experiments & 2 \\ 
    0 & Have software professionals and contribute their effort from your organization (pledge agreements) & 2 \\ \bottomrule
    \end{tabular}
}
\caption{Proposed actions and attendees votes (each attendee present at the end of the meeting was allowed to vote for 3 options)
  \label{tab:actions}}
\end{table}

\subsection{Next steps}

Given these suggestions and the discussion in the workshop, we recommend that the IRIS-HEP project and the HSF consider the following potential actions, related to training, software, and people.

\subsubsection{Training}
    
\begin{itemize}
    
    \item Repurpose nominal funding from in-person training (travel, lodging, food) to pay for the development of training material. This could be done through a set of limited-duration Visiting Pedagogy Fellowships. Each would be aimed at either an existing need for curricular material (as defined by the HSF/IRIS-HEP training group) or proposed by the fellow, but agreed to by the training group. A fellowship project could create initial content for a module, improve the initial content developed by someone else, or both.
    \begin{itemize}
        \item As a concrete need identified today, the large body of training material being used in upcoming HEP C++ training\footnote{\url{https://indico.cern.ch/event/946584/}} could be converted into a more sustainable Carpentries style.
        \item A second example is found in the 2020 US-ATLAS Computing Bootcamp \footnote{\url{https://indico.cern.ch/event/933434/}}, held virtually with 44 participants during August 2020.
        The entire bootcamp had closed captions professionally provided by White Coat Captioning, which facilitated both a deaf bootcamp organizer and deaf participant to fully engage in the bootcamp, and was additionally used by other students to help follow the presented material. All of the material created and taught at the bootcamp is publicly available\footnote{\url{ https://matthewfeickert.github.io/usatlas-computing-bootcamp-2020/}}, and some of the modules taught at the 2020 bootcamp have been made into official HSF training modules\footnote{\url{https://github.com/hsf-training}}.
    \end{itemize}

    \item Invest in scalability of training.
    IRIS-HEP and HSF are collaborating on software training, setting up a model of training across HEP. A framework of a scalable and sustainable software training model has been established that is still its initial phase. Hundreds of people have been trained at several software training and outreach events. The training framework elements\footnote{\url {https://hepsoftwarefoundation.org/workinggroups/training.html}} are:
    \begin{itemize}
            \item publicly shareable software training material  \vspace{-8pt}
            \item a community of trainers \vspace{-8pt}
            \item feedback surveys on usefulness of training \vspace{-8pt}
            \item improvement and funding to sustain training
    \end{itemize}
    Scalability means maximising the impact of this work with the least involvement. Two main factors for scalability are human resources and costs. To self-sustain, scale, and survive means having the ability to train with minimal funding and without direct involvement by HSF or IRIS-HEP. While continuing the existing work, and developing and evolving the training material, we recommend that next major step be spreading the training events and training experts geographically to keep the costs low, and mostly moving to an online training model,  reducing in-person training. Training should be structured such that only a minimal number of people are needed to keep the training infrastructure running and identify what are additional costs for additional events. We need a funding model beyond IRIS-HEP. Mentoring the trainers/mentors to increase the community is an important aspect of sustaining the workforce. In addition, giving them recognition can keep the community vibrant, motivated and help in careers. People should continue to see value in our training and how it can advance our field.
    
\end{itemize}

\subsubsection{Software}
    
\begin{itemize}
    
    \item Explicitly consider and invest in tools that enable and support software sustainability. A concrete need is to renew the HSF's template for C++ projects\footnote{\url{https://github.com/HSF/tools}}, moving to a cookiecutter\footnote{\url{https://cookiecutter.readthedocs.io/}} design and revamping its CMake template with modern best practice. Likewise, another need is to work with Scikit-HEP to develop a similar project template for Python modules\footnote{\url{https://scikit-hep.org/}}. These projects would be intended to support the HP community generally, including IRIS-HEP, so that IRIS-HEP would contribute to the HSF templates, rather than creating its own. Finally, the HEP community could contribute to Projects Carpentry\footnote{\url{https://github.com/carpentries-incubator/proposals/issues/2}} lesson development, which is currently under discussion.
    
    \item Increasingly code, and to a lesser extent data, is being shared via containers such as through Docker and Singularity. Containers bring the convenience of execution environment and portability but also challenges in terms of sustainability. IRIS-HEP could play a role in developing best practices and technical solutions that support the use of containers for the sharing and transfer of knowledge and code (e.g., between students to evolve an analysis.) For example, through creation of a `sharing technologies forum', and bringing in notebooks and model sharing that leverage the kubernetes-based infrastructure of the IRIS-HEP Scalable Systems Laboratory and OSG-LHC. The ongoing Snowmass process in the US could provide one of several avenues for a testbed for supporting how people share knowledge, code and data. 

    \item While issues with sustainability of software and related personnel are well identified, Snowmass 2021 provides an opportunity to have these issues heard, especially at the centers of powers and funding. We submitted a letter of interest\footnote{\url{https://www.snowmass21.org/docs/files/summaries/CommF/SNOWMASS21-CommF0_CommF4-CompF0_CompF7_DanielSKatz-038.pdf}} on this topic to start this conversation.
    
    \item Organize a workshop exploring common software and services across HEP, perhaps associated with a major conference such as ACAT or CHEP. Focus on a particular topic (e.g., small matrix linear algebra libraries) could increase the chances of a successful outcome here. 
    
\end{itemize}

\subsubsection{People}
    
\begin{itemize}
    
    \item Organize a workshop exploring career opportunities within and outside HEP, as key to sustainability of HEP software is sustainability of its personnel. To make this productive, department chairs and laboratory management would need to attend.
    
    \item Advocate to establish rewards for software contribution and innovation (similar to how CPAD\footnote{\url{https://www.anl.gov/hep/coordinating-panel-for-advanced-detectors}} does this for the HEP hardware).
    Many experiments (like CMS and ATLAS) are already recognizing software contributions of young scientists by giving awards. EPS already offers a few HEP prizes, and in 2019 Josh Bendavid won the Young Physicist prize and was cited, amongst other things, for `software development'. The list of APS awards does not seem to have such an award. \footnote{\url{https://www.aps.org/programs/honors/listings.cfm}} What we need and recommend is not one award but a few number of them for software contributions and even software training recognition. These rewards will be a good way to keep the talent motivated.
    
\end{itemize}

\section*{Acknowledgments}

We thank the attendees for their active participation in the workshop. We also thank IRIS-HEP and HSF for the opportunity to gather to discuss these issues and suggest paths forward. This workshop was partially supported through the U.S. National Science Foundation (NSF) under Cooperative Agreement OAC-1836650.

\bibliography{references}

\begin{thebibliography}{}

\bibitem[Albrand et~al., 2005]{Albrand}
Albrand, S., Fulachier, J., Collot, J., and Lambert, F. (2005).
\newblock {The TAG Collector: A Tool for ATLAS Code Release Management}.
\newblock In {\em Proceedings of Computing in High Energy Physics and Nuclear
  Physics 2004}.
\newblock \url{https://doi.org/10.5170/CERN-2005-002.531}.

\bibitem[{Astropy Collaboration} et~al., 2018]{astropy:2018}
{Astropy Collaboration} et~al. (2018).
\newblock {The Astropy Project: Building an Open-science Project and Status of
  the v2.0 Core Package}.
\newblock {\em Astronomical Journal}, 156(3):123.
\newblock \url{https://doi.org/10.3847/1538-3881/aabc4f}.

\bibitem[{Astropy Collaboration} et~al., 2013]{astropy:2013}
{Astropy Collaboration}, {Robitaille}, T.~P., {Tollerud}, E.~J., {Greenfield},
  P., {Droettboom}, M., {Bray}, E., {Aldcroft}, T., {Davis}, M., {Ginsburg},
  A., {Price-Whelan}, A.~M., {Kerzendorf}, W.~E., {Conley}, A., {Crighton}, N.,
  {Barbary}, K., {Muna}, D., {Ferguson}, H., {Grollier}, F., {Parikh}, M.~M.,
  {Nair}, P.~H., {Unther}, H.~M., {Deil}, C., {Woillez}, J., {Conseil}, S.,
  {Kramer}, R., {Turner}, J. E.~H., {Singer}, L., {Fox}, R., {Weaver}, B.~A.,
  {Zabalza}, V., {Edwards}, Z.~I., {Azalee Bostroem}, K., {Burke}, D.~J.,
  {Casey}, A.~R., {Crawford}, S.~M., {Dencheva}, N., {Ely}, J., {Jenness}, T.,
  {Labrie}, K., {Lim}, P.~L., {Pierfederici}, F., {Pontzen}, A., {Ptak}, A.,
  {Refsdal}, B., {Servillat}, M., and {Streicher}, O. (2013).
\newblock {Astropy: A community Python package for astronomy}.
\newblock {\em Astronomy and Astrophysics}, 558:A33.
\newblock \url{https://doi.org/10.1051/0004-6361/201322068}.

\bibitem[{ATLAS Collaboration}, 2019]{athena}
{ATLAS Collaboration} (2019).
\newblock Athena.
\newblock \url{https://doi.org/10.5281/zenodo.3933810}.

\bibitem[Baxter et~al., 2012]{Baxter}
Baxter, R., Chue~Hong, N., Gorissen, D., Hetherington, J., and Todorov, I.
  (2012).
\newblock {The Research Software Engineer}.
\newblock In {\em {Proceedings of Digital Research 2012 Conference (DR12)}}.
\newblock \url{http://purl.org/net/epubs/work/63787}.

\bibitem[Bhandari~Neupane et~al., 2019]{python-os-bug}
Bhandari~Neupane, J., Neupane, R.~P., Luo, Y., Yoshida, W.~Y., Sun, R., and
  Williams, P.~G. (2019).
\newblock {Characterization of Leptazolines A-D, Polar Oxazolines from the
  Cyanobacterium Leptolyngbya sp., Reveals a Glitch with the "Willoughby-Hoye"
  Scripts for Calculating NMR Chemical Shifts}.
\newblock {\em Organic Letters}, 21(20):8449--8453.
\newblock \url{https://doi.org/10.1021/acs.orglett.9b03216}.

\bibitem[Brett et~al., 2017]{uk-rse-nation}
Brett, A., Croucher, M., Haines, R., Hettrick, S., Hetherington, J., Stillwell,
  M., and Wyatt, C. (2017).
\newblock {Research Software Engineers: State of the Nation}.
\newblock \url{https://doi.org/10.5281/zenodo.495360}.

\bibitem[Brun et~al., 2012]{web-grid}
Brun, R., Carminati, F., and Carminati, G.~G., editors (2012).
\newblock {\em From the WEB to the GRID and Beyond, Computing Paradigms Driven
  by High-Energy Physics}.
\newblock Springer.
\newblock \url{https://doi.org/10.1007/978-3-642-23157-5}.

\bibitem[Cabunoc~Mayes, 2020]{Mayes_2020}
Cabunoc~Mayes, A. (2020).
\newblock Work open, lead open.
\newblock Chan-Zuckerberg Initiative (CZI) Essential Open Source Software
  (EOSS) Kickoff Meeting, Berkeley, California, USA.

\bibitem[Couturier et~al., 2017]{couturier2017hep}
Couturier, B., Eulisse, G., Grasland, H., Hegner, B., Jouvin, M., Kane, M.,
  Katz, D.~S., Kuhr, T., Lange, D., Lorenzo, P.~M., Ritter, M., Stewart, G.~A.,
  and Valassi, A. (2017).
\newblock {HEP} software foundation community white paper working group -
  software development, deployment and validation.
\newblock \url{https://arxiv.org/abs/1712.07959}.

\bibitem[Crouch et~al., 2013]{Crouch_2013}
Crouch, S., Chue~Hong, N.~P., Hettrick, S., Jackson, M., Pawlik, A., Sufi, S.,
  Carr, L., De~Roure, D., Goble, C., and Parsons, M. (2013).
\newblock {The Software Sustainability Institute: Changing Research Software
  Attitudes and Practices}.
\newblock {\em Computing in Science \& Engineering}, 15(6).
\newblock \url{https://doi.org/10.1109/MCSE.2013.133}.

\bibitem[Elmer et~al., 2017]{Elmer:2017rej}
Elmer, P., Neubauer, M., and Sokoloff, M.~D. (2017).
\newblock {Strategic Plan for a Scientific Software Innovation Institute
  ({S2I2}) for High Energy Physics}.
\newblock \url{https://arxiv.org/abs/1712.06592}.

\bibitem[Gamblin et~al., 2015]{7832814}
Gamblin, T., LeGendre, M., Collette, M.~R., Lee, G.~L., Moody, A., de~Supinski,
  B.~R., and Futral, S. (2015).
\newblock The {Spack} package manager: bringing order to {HPC} software chaos.
\newblock In {\em SC15: International Conference for High-Performance
  Computing, Networking, Storage and Analysis}, pages 1--12, Los Alamitos, CA,
  USA. IEEE Computer Society.
\newblock \url{https://doi.ieeecomputersociety.org/10.1145/2807591.2807623}.

\bibitem[Hegner et~al., 2020]{hegner_benedikt_2020_3965581}
Hegner, B., Morgan, B., and Stewart, G.~A. (2020).
\newblock {\em Proposal for HSF Project Best Practices}.
\newblock \url{https://doi.org/10.5281/zenodo.3965581}.

\bibitem[Hettrick, 2018]{2014_Research_Software_Survey}
Hettrick, S. (2018).
\newblock {UK Research Software Survey 2014}.
\newblock \url{https://doi.org/10.5281/zenodo.1183562}.

\bibitem[Hinsen, 2019]{Hinsen_2019}
Hinsen, K. (2019).
\newblock Dealing with software collapse.
\newblock {\em Computing in Science \& Engineering}, 21(3):104–--108.
\newblock \url{https://doi.org/10.1109/MCSE.2019.2900945}.

\bibitem[Jouvin et~al., 2016]{jouvin_m_2016_1469636}
Jouvin, M., Harvey, J., McNab, A., Sexton-Kennedy, E., and Wenaus, T. (2016).
\newblock {\em Software Licence Agreements HSF Policy Guidelines}.
\newblock \url{https://doi.org/10.5281/zenodo.1469636}.

\bibitem[Labra et~al., 2020]{phoenix}
Labra, E.~C., Ali, F., Moyse, E., Couturier, B., and Bianchi, R.~M. (2020).
\newblock Hsf/phoenix: Interim phoenix version.
\newblock \url{https://doi.org/10.5281/zenodo.3925404}.

\bibitem[Merali, 2010]{Merali-sci-computing}
Merali, Z. (2010).
\newblock {Computational Science: ... Error ... Why Scientific Programming Does
  Not Compute)}.
\newblock {\em Nature}, 467:775--777.
\newblock \url{https://doi.org/10.1038/467775a}.

\bibitem[Miller, 2006]{science-retraction}
Miller, G. (2006).
\newblock {A Scientist’s Nightmare: Software Problem Leads to Five
  Retractions}.
\newblock {\em Science}, 314(5807):1856--1857.
\newblock \url{10.1126/science.314.5807.1856}.

\bibitem[Philippe et~al., 2019]{2018_International_RSE_Survey}
Philippe, O., Hammitzsch, M., Janosch, S., van~der Walt, A., van Werkhoven, B.,
  Hettrick, S., Katz, D.~S., Leinweber, K., Gesing, S., Druskat, S., Henwood,
  S., May, N.~R., Lohani, N.~P., and Sinha, M. (2019).
\newblock softwaresaved/international-survey: Public release for 2018 results
  (version 2018-v.1.0.2).
\newblock \url{http://doi.org/10.5281/zenodo.2585783}.

\bibitem[Porcelli, 2013]{Porcelli_2013}
Porcelli, J. (2013).
\newblock How to grow users into active community members and get your
  community more engaged.
\newblock 2013 Open Source Software Summit, Washington, DC, USA.

\bibitem[Salzburger et~al., 2020]{acts}
Salzburger, A., Schlag, B., Gumpert, C., Klimpel, F., Grasland, H., Hrdinka,
  J., Kiehn, M., Calace, N., Gessinger, P., Langenberg, R., and Ai, X. (2020).
\newblock Acts project: v0.20.00.
\newblock \url{https://doi.org/10.5281/zenodo.3741401}.

\bibitem[Schreiner, 2020]{ModernCMake}
Schreiner, H. (2020).
\newblock {Modern CMake}.
\newblock \url{https://gitlab.com/CLIUtils/modern-cmake}.

\bibitem[Sufi and Jay, 2018]{Sufi_2018}
Sufi, S. and Jay, C. (2018).
\newblock Raising the status of software in research: A survey-based evaluation
  of the software sustainability institute fellowship programme [version 1;
  peer review: 3 approved with reservations].
\newblock {\em F1000Research}, 7(1599).
\newblock \url{https://doi.org/10.12688/f1000research.16231.1}.

\bibitem[{The HEP Software Foundation} et~al., 2019]{Albrecht2019}
{The HEP Software Foundation} et~al. (2019).
\newblock A roadmap for {HEP} software and computing {R{\&}D} for the 2020s.
\newblock {\em Computing and Software for Big Science}, 3(1):7.
\newblock \url{https://doi.org/10.1007/s41781-018-0018-8}.

\end{thebibliography}

\end{document}